\documentstyle[epsf,eqsecnum,floats,preprint,aps]{revtex}

\tighten

\input epsf
\begin{document}

\newcommand{\be}{\begin{equation}}
\newcommand{\ee}{\end{equation}}
\newcommand{\bea}{\begin{eqnarray}}
\newcommand{\eea}{\end{eqnarray}}
\newcommand{\PSbox}[3]{\mbox{\rule{0in}{#3}\includegraphics{#1}\hspace{#2}}}
\newcommand{\modified}[1]{{\it #1}}
\newcommand{\eos}{{\rm w}}
\newcommand{\eff}{{\rm eff}}

\def\5M{M^3_{(5)}}
\def\4M{M^2_{(4)}}
\def\mpsq{M_{\rm P}^2}
\def\om{\Omega_m}
\def\omt{\Omega_m^0}

\overfullrule=0pt
\def\Int{\int_{r_H}^\infty}
\def\d{\partial}
\def\e{\epsilon}
\def\M{{\cal M}}
\def\high{\vphantom{\Biggl(}\displaystyle}
\catcode`@=11
\def\@versim#1#2{\lower.7\p@\vbox{\baselineskip\z@skip\lineskip-.5\p@
    \ialign{$\m@th#1\hfil##\hfil$\crcr#2\crcr\sim\crcr}}}
\def\simge{\mathrel{\mathpalette\@versim>}} % 
\def\simle{\mathrel{\mathpalette\@versim<}} % 
\def\sun{\hbox{$\odot$}}
\catcode`@=12 % at signs are no longer letters

%\rightline{CWRU--PXX--XX}
%\rightline{CERN--PH--TH/2004--XXX}
\rightline{astro-ph/0408246}
\vskip 4cm

\setcounter{footnote}{0}

\begin{center}
\large{\bf How a brane cosmological constant can trick us into thinking that $\eos < -1$}
\ \\
\ \\
\normalsize{Arthur Lue\footnote{E-mail: lue@cern.ch}
  and Glenn D. Starkman\footnote{E-mail: glenn.starkman@cern.ch}}
\ \\
\ \\
\small{\em
Center for Education and Research in Cosmology and Astrophysics\\
Department of Physics, Case Western Reserve University,
Cleveland, OH 44106--7079\\
and CERN Physics Department, Theory Division, CH--1211 Geneva 23, Switzerland}

\end{center}

\begin{abstract}
\noindent
Observations exploring the contemporary cosmic acceleration have
sparked interest in dark energy models possessing equations of state
with $\eos < -1$.  We review how the cosmic expansion history of a
Dvali--Gabadadze--Porrati (DGP) braneworld model with a standard
brane cosmological constant can mimic that of ordinary 4--dimensional
gravity with $\eos<-1$ ``phantom" dark energy for observationally
relevant redshifts.  We reinterpret the effective phantom nature of the
dark energy as arising from dynamical-screening of the brane
cosmological constant in DGP.  This unusual variety of expansion
history is thus possible without violating the null-energy condition,
without ghosts and without any big rip, in a model which seems no
more contrived than most evolving dark energy models.  We indicate
ways by which one may observationally test this effective $\eos < -1$
possibility, and differentiate it from ``ordinary'' phantom dark-energy.
\end{abstract}

\setcounter{page}{0}
\thispagestyle{empty}
\maketitle

\eject

\vfill

\baselineskip 18pt plus 2pt minus 2pt

\section{Introduction}

The discovery of a contemporary cosmic
acceleration~\cite{Perlmutter:1998np,Riess:1998cb} is one of the most
profound scientific observations of the 20th century.  The conventional
explanation is some unknown smooth dark energy component that
possesses a sufficiently negative pressure.  Then, the Friedmann equation
for a spatially flat Universe becomes
\be
	H^2 = {8\pi G\over 3}\left(\rho_M + \rho_{DE}\right)\ ,
\label{oldFried}
\ee
the dark energy has an equation of state $\eos$, where
\be
      \rho_{DE}(t) = \rho^0_{DE}a^{-3(1+\eos)}\ .
\ee

The quantity $a(t)$ is the scale factor of the Universe.  Dark
energy composed of just a cosmological constant ($\eos = -1$) is
fully consistent with existing observational data.  However, that
same data also suggests that if one is guided solely by the
expansion history of the universe, $w <-1$ remains a viable
possibility, and is slightly favored \cite{Riess}
(though with little statistical significance).

Regardless of the strength of this inference, $\eos<-1$
remains a logical and intriguing possibility.  
Many attempts have been made to find dark energy models that allow 
so-called phantom dark energy:  dark energy with $w < -1$
\cite{Caldwell:1999ew,Barreiro:1999zs,Chiba:1999ka,Onemli:2002hr,Carroll:2003st,Melchiorri:2002ux,Cline:2003gs,Onemli:2004mb}.   
Recently, Carroll, De~Felice and Trodden \cite{BD} articulated
the possibility that Friedmann equation Eq.~(\ref{oldFried}) does not describe
nature and that one may obtain a $\eos_{\eff} < -1$ through a modification of
the Friedmann equation rather than through an unusual (and often pathological)
form of dark energy.  They restricted their approach to Brans--Dicke--type
theories and found ``that only highly contrived models would lead observers to measure $\eos < -1$.''

Nevertheless, it is already known how one may achieve this goal within
the context of modified-gravity theories.  Several years ago, 
Sahni and Shtanov pointed out how to reproduce $\eos_{\eff} < -1$ cosmologies \cite{Sahni:2002dx} in 
the well-studied Dvali, Gabadadze and Porrati (DGP) braneworld theory \cite{Dvali:2000hr,Dvali:2001gm,Dvali:2001gx}
using the governing cosmological equations first developed by Deffayet \cite{Deffayet}.
This braneworld model is a complete theory,
contains only matter satisfying the null-energy condition, 
exhibits no ghosts in the classical field theory, 
avoids a future big rip 
and requires no contrived system parameters.\footnote{
	There is currently a debate regarding {\em quantum} strong coupling
	in this theory \cite{Luty:2003vm}.  While this subject remains an important
	issue to resolve, it has no bearing on the present context where we are
	solely concerned with classical field evolution.}
In this note, we review the features of this braneworld model, taking a
somewhat different approach for instructive purposes.  
We then discuss signatures by which one may identify this $\eos_{\eff} < -1$ possibility 
observationally and distinguish it from ``ordinary'' phantom dark energy.

\section{Screening of the cosmological constant}

The DGP model is a 5-dimensional  braneworld with modified
gravity and thus a modified Friedmann equation \cite{Deffayet}.  
Until now, phenomenological emphasis has been put on the possibility in this model of 
``self-acceleration'' -- acceleration with {\em no} negative-pressure energy-momentum component.  
Observationally, this leads to $\eos_{\eff} > -1$, 
and this self-accelerating model has been studied extensively
\cite{Deffayet,Deffayet:2001pu,Deffayet:2002sp,Alcaniz:2002qh,Lue:2002sw,Dvali:2002vf,Lue:2004rj}.
However, there is another cosmological phase besides the
self-acceleration one \cite{Deffayet}.  If one allows the most trivial
negative-pressure energy-momentum component (a cosmological constant)
then one can have $\eos_{\eff} < -1$ \cite{Sahni:2002dx}.  We will rely on
well-known governing equations already set out in the DGP literature, and will
liberally appeal to that literature for details.

We invoke the following scenario:  ordinary particles and fields, other than the
graviton, are confined to a three-dimensional brane hypersurface
embedded in a five-dimensional, infinite-volume Minkowski bulk \cite{Dvali:2000hr}. 
The bulk is empty; all energy-momentum is isolated on the brane.  
The graviton is metastable on the brane.  
The action is \cite{Dvali:2000hr}:
\be
S_{(5)} = -\frac{1}{16\pi}M^3 \int d^5x
\sqrt{-g}~R +\int d^4x \sqrt{-g^{(4)}}~{\cal L}_m \ .
\label{action}
\ee
$M$ is the fundamental five-dimensional Planck scale.
The first term in $S_{(5)}$ is the 
Einstein-Hilbert action in five dimensions for a five-dimensional
metric $g_{AB}$ (bulk metric) with Ricci scalar $R$ and determinant $g$.  
The metric $g^{(4)}_{\mu\nu}$ is the induced (four-dimensional) metric on
the brane, and $g^{(4)}$ is its determinant.
An intrinsic curvature term is added to the brane
action~\cite{Dvali:2000hr}:
\be
-\frac{1}{16\pi}M^2_P \int d^4x \sqrt{-g^{(4)}}\ R^{(4)}\ .
\label{action2}
\ee
Here, $M_P$ is the observed four-dimensional Planck scale
(see~\cite{Dvali:2000hr,Dvali:2001gm,Dvali:2001gx} for
details). Equation~(\ref{action2}) is the Einstein-Hilbert
action for the induced brane metric $g^{(4)}_{\mu\nu}$,
$R^{(4)}$ being its scalar curvature.    

Gravity in this model appears four-dimensional Einstein at short distances, 
while at larger distances, gravity evaporates into the bulk and is able to ``probe''
the extra dimension.  The scale that governs that crossover is
\be
	r_0 = {\mpsq \over 2M^3}\ .
\label{r0}
\ee
The cosmology on the brane for a spatially-flat brane Universe follows
Deffayet's modified Friedmann equation \cite{Deffayet}
\be
     H^2 \pm {H\over r_0} = {8\pi G\over 3}\rho(t)\ ,
\label{Fried}
\ee
where $H(t) = \dot{a}/a$ is the Hubble parameter.  The quantity
$\rho(t)$ is the density of all energy momentum on the brane.  The
two choices of sign represent two distinct cosmological phases.  The
celebrated self-accelerating phase
\cite{Deffayet,Deffayet:2001pu,Deffayet:2002sp,Alcaniz:2002qh}
corresponds to the lower sign, while we refer to the upper sign
as the Minkowski cosmological phase.  A natural geometric
interpretation exists for both phases \cite{Deffayet,Lue:2002fe}, but
here we focus on the Minkowski phase.  The key to this phase
is that the brane is extrinsically curved in such a way that shortcuts
through the bulk allow gravity to screen the effects of the
brane energy-momentum contents at Hubble parameters $H \sim r_0^{-1}$.
This is not the case for the self-accelerating phase.

Take the observed Universe (which is our brane) to have a homogeneous
distribution of pressureless matter as well as a cosmological constant.
Then,
\be
     H^2 + {H\over r_0} = {8\pi G\over 3}\rho_M(t) + \lambda\ .
\label{Fried2}
\ee
One may infer an effective dark energy component,
c.f. Eq.~(\ref{oldFried}), of the form
\be
     {8\pi G\over 3}\rho^{\eff}_{DE} = \lambda - {H\over r_0}\ .
\label{DEeff}
\ee
Since $H(t)$ is a decreasing function of time, this effective dark
energy component is increasing with time, the hallmark of a $\eos_{\rm
eff} < -1$ component.  Indeed, one can easily define a $\eos_{\eff}$
using Eq.~(\ref{oldFried}) and see that during the epoch when dark
energy dominates, $\eos_{\eff} < -1$.  The second term in Eq.~(\ref{DEeff})
is the effect of gravitational screening of the cosmological constant.
As the universe evolves, that screening becomes increasingly ineffective
and the effective dark-energy density appears to {\em increase}.  
The degree to which this effective
dark-energy component grows with time is modulated by $r_0$ and in
the limit $r_0 \rightarrow \infty$, we recover the standard $\Lambda$CDM cosmology.  For any value of $r_0$, $\eos_{\eff} \rightarrow -1$ with time
and a pure cosmological-constant Universe is restored in the distant future.\footnote{
	In the past, when $\rho_M$ dominates, the definition for $\eos_{\eff}$
	breaks down, particularly when $\rho^{\eff}_{DE} = 0$.  This anomaly
	is an artifact of the definition: the evolution of $H(t)$ remains
	well-behaved throughout time.}

\section{Expansion history}

Equations~(\ref{Fried2}) and (\ref{DEeff}) imply that the cosmic expansion
history does not strictly follow that for a conventional constant-$\eos$
dark energy model.  The precise time evolution of the Hubble parameter for
this screened-cosmological-constant DGP may be determined explicitly from Eq.~(\ref{Fried2}).  
Rewriting $H(t)$ in terms of more phenomenological quantities,
we have
\be
     {H(z)\over H_0} = {1\over 2}\left[-{1\over r_0H_0}
       + \sqrt{\left(2+{1\over r_0H_0}\right)^2
	 + 4\Omega_M^0\left((1+z)^3-1\right)}\right]\ ,
\label{hubble}
\ee
where the redshift, $z$, today's Hubble parameter, $H_0$, and
$\Omega^0_M = 8\pi G\rho^0_M/3H_0^2$ are all defined in the usual way.
Given $H_0$, there are two dimensionless parameters that may be adjusted
phenomenologically: $\Omega^0_M$ and $r_0H_0$.  This is similar to the conventional
dark-energy picture, Eq.~(\ref{oldFried}), which also has two phenomenological
parameters: $\Omega_M^0$ and $\eos$.  

\begin{figure}[t]
\centerline{\epsfxsize=5cm\epsffile{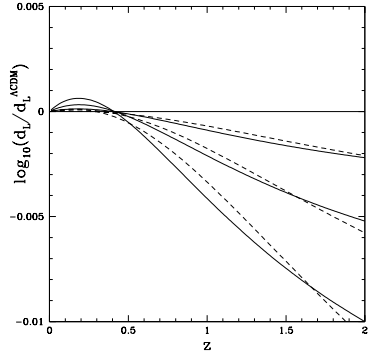}}
\caption{
$d^\eos_L(z)/d^{\Lambda CDM}_L(z)$ and $d^{\lambda DGP}_L(z)/d^{\Lambda
CDM}_L(z)$ for a variety of models.  The reference model is the
best-fit flat $\Lambda$CDM, with $\Omega^0_M =0.27$.  The dotted
curves are for the constant-$\eos$ models with $(\Omega^\eos_M,\eos) =
(0.28,-1.02)$,$(0.294,-1.05)$, and $(0.316,-1.1)$ from top to bottom.
The quantity $\Omega^\eos_M$ is chosen to match the $\Lambda$CDM model at
low--$z$ for a given $\eos$.  The solid curves are for the
screened-cosmological-constant DGP models with $(\Omega^{\lambda DGP}_M,r_0H_0) =
(0.292,5.9)$,$(0.34,1.83)$, and $(0.42,0.86)$ from top to bottom where
the parameters were chosen to best mimic the corresponding
$\eos$--models.
}
\label{fig:d}
\end{figure}

When comparing the expansion histories $a(t)$ of specific examples of
screened-cosmological-constant DGP models ($\lambda$DGP) with
examples of constant-$\eos$ dark-energy models, we refer to the
$\Omega_M$--parameter of the first as $\Omega^{\lambda DGP}_M$
and the second as $\Omega^\eos_M$.  For models with closely corresponding
histories, these will not be the same.  The luminosity-distance takes the
standard form in spatially flat cosmologies:
\be
      d^{\lambda DGP}_L(z) = (1+z)\int_0^z {dz\over H(z)}\ ,
\label{dDGP}
\ee
using Eq.~(\ref{hubble}).  In Fig.~\ref{fig:d}, we compare this distance
with the luminosity distance for a constant-$\eos$ dark-energy model
\be
     d^\eos_L(z) = (1+z)\int_0^z {H_0^{-1}dz\over \sqrt{\Omega^\eos_M(1+z)^3
	 + (1-\Omega^\eos_M)(1+z)^{3(1+\eos)}}}\ ,
\label{dQ}
\ee
both normalized to the best-fit $\Lambda$CDM model.
Figure~\ref{fig:d} demonstrates that while the expansion histories of
the currently considered model and constant-$\eos$ dark-energy
models are not identical, they are for all practical purposes
indistinguishable using luminosity-distance data.  One should not put
too much weight onto the particular values of $\Omega_M^{\lambda DGP}$
in Fig.~\ref{fig:d} as we should be more interested in having this
$\lambda$DGP model best match the data, rather than some $\eos$
dark-energy cosmology.  For a more thorough analysis of best-fit
parameters of this model, see Alam and Sahni \cite{Alam:2002dv}.

\section{Observational Signatures}

Up until now, we have reviewed what exists in the literature (though
presented in a slightly different way) but, for whatever reason, remains
largely unappreciated by the cosmology community.  However, we can
now ask important questions about observational constraints of this
$\lambda$DGP model, and in particular how one may differentiate this
modified-gravity model from dark-energy models that may mimic the same 
cosmic expansion history.

The major tests of this theory come from constraints on the modification
of the gravitational force law.  Because this model is in the class of more
general DGP models, it does not suffer from the same constraints as
found for Brans--Dicke theories of gravity.  Extra scalar degrees of freedom
are suppressed at distance scales much shorter than $r_0 \sim H_0^{-1}$,
thus we expect current observational constraints on the form of
gravity are not spoiled (see the discussion in \cite{Deffayet:2001uk,Lue:2001gc,Gruzinov:2001hp,Porrati:2002cp,Lue:2002sw,Dvali:2002vf}).

Generalizing the analysis found in Refs.~\cite{Lue:2002sw,Lue:2004rj}
we may ascertain the modification of the Schwarzschild metric for a
spherically symmetric source for any background cosmology.  The
metric on the brane
\be
	ds^2 = g_{00}dt^2 - g_{rr}dr^2 - r^2 d\Omega\ .
\ee
For a cosmological background with {\em arbitrary} evolution $H(t)$,
we find that
\bea
rg'_{00}(t,r)|_{\rm brane} &=& {R_g\over r}\left[1+\Delta(r)\right] - 2(H^2+\dot{H})r^2
\label{brane-n}\\
g_{rr}(t,r)|_{\rm brane} &=& 1 + {R_g\over r}\left[1-\Delta(r)\right] + H^2r^2\ ,
\label{brane-a}
\eea
where dot now denotes differentiation with respect to $t$ and prime
denotes differentiation with respect to $r$.  Note that the cosmological
background contribution is included in these metric components.  The
quantity $R_g$ is defined as
\be
	R_g(r) = 8\pi G \int_0^r dr r^2 ~\delta\rho(r)\ ,
\ee
where $\delta\rho(r)$ is density of matter in excess of the background
cosmological density;  $\Delta(r)$ is defined as
\be
     \Delta(r) = {3\beta r^3\over 4 r_0^2R_g}
     \left[\sqrt{1+{8r_0^2R_g\over 9\beta^2r^3}}-1\right]\ ;
\label{Delta}
\ee
and
\be
     \beta = 1\pm2r_0H\left(1 + {\dot{H}\over 3H^2}\right)\ .
\label{beta}
\ee
Just as for the modified Friedmann equation, Eq.~(\ref{Fried}), 
there is a sign degeneracy in Eq.~(\ref{beta}).  
The lower sign corresponds to the self-accelerating
cosmologies.  For the cosmological phase in which we are interested 
($\lambda$DGP), we take the upper (positive) sign.\footnote{
	There may be evidence of alternative nonperturbative
	phases of compact matter sources \cite{Gabadadze:2004iy}.}

\subsection{Solar system}

For solar system measurements, one may take a limit in
Eqs.~(\ref{brane-n}) and (\ref{brane-a}) where $r^3 \ll r_0^2R_g$.
Then, there are only small corrections to the usual Schwarzschild
metric.  To leading-order,
the immediate consequence of this modification of the Schwarzschild
metric is that all orbiting bodies suffer a universal anomalous precession
of \cite{Lue:2002sw}
\be
     {d\over dt}\Delta\phi = -{3\over 8r_0}\ .
\ee
If $r_0H_0 = 1$, then the precession rate is about 3~$\mu{\rm as/year}$.
Note that this rate is independent of the source mass and only depends
on the distance scale $r_0$ which characterizes the scale at which
bulk effect manifest themselves in the Universe.
Current solar system constraints are around 10~$\mu{\rm as/year}$, and improvements can be
expected over the next decade \cite{Lue:2002sw,Dvali:2002vf}.  It is
intriguing that one may gain insight into physics on the largest of
cosmological scales from local tests of gravity within our solar system.

\subsection{Large scale structure}

One may now also invoke tests from large scale structure.  For linear
perturbations, Eqs.~(\ref{brane-n}) and (\ref{brane-a}) yield a metric
for a linearized scalar-tensor theory with Brans--Dicke parameter
\be
	\omega = {3\over 2}(\beta - 1)\ .
\ee
Note this is only true for weak, linear perturbations around the cosmological
background, and that for solar-system tests one must refer to the previous
subsection.  However, for the linear growth of structure, this is perfectly
adequate.  Nonrelativistic matter perturbations with
$\delta(t) = \delta\rho/\rho$ evolve via
\be
	\ddot{\delta} + 2H\dot{\delta} = 4\pi G\rho\left(1+{1\over 3\beta}\right)\delta\ ,
\label{deltaevol}
\ee
implying the effect of $\lambda$DGP on linear matter pertubations is
that self-attraction of overdensities is governed by an evolving $G_{\eff}$:
\be
	G_{\eff} = G\left[1 + {1\over 3\beta}\right]\ .
\label{Geff}
\ee
This effective Newton's constant runs from $G$ at early
times to values larger than $G$ as redshift $z \rightarrow 0$.
This mild strengthening of the self-attraction of density perturbations
will increase perturbation growth, even beyond that one would expect
from that from a dark energy model that mimics the same cosmic
expansion history.

For example, by numerically evolving Eq.~(\ref{deltaevol}), one finds
for the corresponding parameters $(\Omega^{\lambda DGP}_M,r_0H_0)
= (0.3,3.77), (0.35,1.41)$ and $(0.4,0.870)$, we get anomalously high
growth of $\lambda$DGP (beyond that one would get for a dark-energy
model with the same cosmic expansion history) of $1.5\%$,
$4.6\%$ and $9.3\%$, respectively.  This discrepancy of linear
growth is roughly proportional to $1 - \eos_{\eff}$ and provides an
intriguing handle by which to differentiate between this modified-gravity
scenario and corresponding dark-energy ones.

\section{Concluding remarks}

We illustrate in this note how one may mimic a $\eos<-1$ dark-energy
cosmology using just a brane cosmological constant in DGP in the
Minkowski cosmological phase.  The effect of this braneworld scenario
is that leakage of gravity off the brane at late times screens the
effect of the cosmological constant.  This screening attenuates with
time, yielding a dark-energy density that appears to grow as the
universe evolves, thus providing a cosmic evolution with $\eos_{\eff}<-1$.
There are no currently catastrophic observational constraints on this
model, though data in the near future may provide some intriguing tests. 
There may be an important roll for modified gravity to play
in understanding the details of what drives today's cosmic acceleration.

\begin{acknowledgments}
The authors thank G.~Gabadadze and R.~Scoccimarro for helpful communications.  
We thank the CERN Theory Division for its hospitality.
This work is sponsored by DOE Grant DEFG0295ER40898, 
the CWRU Office of the Provost, and CERN.  
GDS thanks Maplesoft for the use of Maple V.
\end{acknowledgments}

\end{document}